\documentclass[11pt,a4paper]{article}
\usepackage[T1]{fontenc}
\usepackage{lmodern}
\usepackage[margin=20mm]{geometry}
\usepackage{amsmath,amssymb,graphicx,array}
\usepackage[table]{xcolor}
\usepackage[hidelinks]{hyperref}
\usepackage{cite}
\title{Single-sided lens terahertz flat-top beam shaping in the Fresnel region}

\author{Ehsan Imanbeygi$^{1}$, Mohammad Hossein Koohi Ghamsari$^{1}$,\\
Mehdi Ahmadi-Boroujeni$^{1,*}$, and Babak Honarbakhsh$^{2}$\\
\small $^{1}$Department of Electrical Engineering, Sharif University of Technology, Tehran, Iran\\
\small $^{2}$Faculty of Electrical Engineering, Shahid Beheshti University, Tehran, Iran\\
\small $^{*}$Corresponding author: ahmadi@sharif.edu}
\date{}

\begin{document}
\maketitle

\noindent\textbf{Author-revised manuscript.} Published in \emph{Applied Optics}, \textbf{65}(2), 551--557 (2026).\par
\noindent\href{https://doi.org/10.1364/AO.583170}{doi:10.1364/AO.583170}.

\medskip
\noindent{\footnotesize \textcopyright\ 2026 Optica Publishing Group. One print or electronic copy may be made for personal use only. Systematic reproduction and distribution, duplication of any material in this paper for a fee or for commercial purposes, or modifications of the content of this paper are prohibited. \href{https://opg.optica.org/ao/abstract.cfm?uri=ao-65-2-551}{Online journal abstract.}}

\begin{abstract}
In this paper, a novel analytical approach is introduced for the design of a single-sided plano-convex lens (PCL) capable of terahertz (THz) flat-top beam shaping in the Fresnel region. The lens profile, derived from geometrical optics (GO) principles, integrates collimating and beam-shaping functions into a single optical surface, thereby simplifying fabrication. The efficiency of the method is demonstrated by designing two PCLs that convert a divergent Gaussian beam from a monochromatic THz source (at 100 GHz and 200 GHz) into flat-top beams with target radii of 400 mm and 500 mm at distances of 3 m and 4 m, respectively. The resulting fields, evaluated through analytical, integral equation (IE), and finite element (FEM) techniques, are shown to be comparable to those produced by a conventional double-sided concave-convex lens (CCL). Furthermore, the sensitivity of the design to frequency and alignment variations is investigated, confirming the robustness of this single-curvature approach for integration into practical THz imaging systems.
\end{abstract}


\section{Introduction}

In recent years, significant research focus has been directed toward the millimeter-wave (mm-Wave) and terahertz (THz) frequency spectrum, owing to its broad potential in diverse applications ranging from high-speed wireless communications to advanced sensing and spectroscopy \cite{bauer2022terahertz}. The THz band uniquely bridges the gap between microwave and infrared frequencies, offering a combination of electronic and photonic characteristics that enable high-resolution, non-invasive probing of materials. The suitability of THz radiation for non-destructive inspection and imaging stems from several key physical properties: it is non-ionizing and thus safe for biological and sensitive materials; it can penetrate a wide variety of non-conductive or obscuring substances such as plastics, fabrics, and paper; yet it exhibits strong reflection from metallic surfaces, making it highly advantageous for the analysis of concealed or layered structures \cite{li2023high}. These distinctive attributes have motivated the development of THz imaging systems for applications in security screening, biomedical diagnostics, and material characterization.

In response to these growing demands, extensive research efforts have been devoted to designing optical setups that can efficiently manipulate and image THz radiation in the Fresnel region, where the wavefront curvature plays a crucial role in determining image quality. Both reflective and transmissive architectures have been explored to enhance focusing precision, imaging resolution, and system compactness. Reflective systems such as confocal and off-axis parabolic mirror arrangements have demonstrated excellent efficiency and reduced chromatic aberration \cite{10999938,koohi2025confocal,Kasalynas:13,lai2024single,Xu:24,8995564}, while transmissive systems based on diffractive or refractive elements have enabled lightweight and cost-effective alternatives with promising imaging performance \cite{ning2023binary,hajitabarmarznaki2023wideband,zatta2021silicon,stephen2023terahertz}. The continuous advancement of these optical configurations highlights the importance of optimizing both structural design and electromagnetic performance to achieve accurate and high-quality THz imaging across various application domains.

Beam shaping, the optical counterpart of beamforming in communications, involves tailoring a beam’s spatial intensity distribution to achieve specific illumination profiles and performance objectives across a range of optical and terahertz systems \cite{Chen:25}. One of the primary goals in Fresnel-region imaging is the generation of uniform, flat-top illumination at a target distance, which significantly enhances image contrast, accuracy, and system calibration stability. Achieving such uniformity is critical for applications in active imaging, laser processing, and material inspection, where uneven irradiance can degrade the quality or reliability of the measured data. To address these needs, numerous beam shaping techniques have been developed based on both reflective and transmissive architectures, each offering unique advantages in terms of efficiency, compactness, and fabrication flexibility \cite{imanbeygi2024gaussian,chen2024design,zhang2025terahertz,abbaszadeh2019generating,price2020subterahertz,bezziou2023design,kleindienst2010highly}.

Reflective beam shaping systems have attracted significant interest due to their low chromatic aberration, robustness to misalignment, and suitability for integration into compact setups. For instance, \cite{chen2024design} proposed a dual-surface freeform reflective design that reshapes elliptical Gaussian beams emitted from laser diodes into collimated beams with substantially reduced divergence, thereby improving the uniformity and efficiency of optical sources. Similarly, \cite{imanbeygi2024gaussian} demonstrated a metallic reflectarray operating at 100 GHz capable of transforming a Gaussian horn beam into a high-quality flat-top pattern in an off-axis reflective mode. Their approach combined geometrical-optics-based phase synthesis with full-wave simulations, producing a compact and easily fabricated solution for millimeter-wave imaging. These reflective systems highlight the advantages of precision phase control and structural simplicity, positioning them as promising candidates for next-generation terahertz and millimeter-wave imaging setups.

In parallel, transmissive beam shaping devices have evolved rapidly, driven by the growing need for lightweight, low-cost, and broadband solutions. Classical refractive designs, such as flat-top beam shaper (FBS) lenses, have shown excellent results in converting pseudo-Gaussian beams into sharply defined flat-top profiles with minimal ripple and tunable beam radii \cite{abbaszadeh2019generating}. More recent innovations employ metasurfaces and diffractive metalenses, which enable subwavelength phase manipulation and superior efficiency. Recently, a terahertz metasurface employing an improved Gerchberg–Saxton algorithm is introduced to achieve a wideband flat-top beam with high shaping efficiency and low transmission loss \cite{zhang2025terahertz}, while \cite{Chen:25} demonstrated a metalens-based beam shaper that converts Gaussian beams into circular and square flat-top beams with over 90\% efficiency. Complementary efforts using 3D-printed aspheric and Powell-type lenses have further expanded the design space. For example, references \cite{price2020subterahertz} and \cite{bezziou2023design} fabricated compact polymer-based lenses for sub-terahertz and millimeter-wave imaging, demonstrating excellent uniformity and cost-effectiveness. Similarly, in another work, an efficient THz illumination system was developed, integrating a Gaussian-to-top-hat beam shaper with experimentally validated tolerance-aware design \cite{kleindienst2010highly}. Therefore, these transmissive and hybrid approaches represent a transformative step toward compact, efficient, and highly uniform illumination systems tailored for modern terahertz and optical imaging applications.

In this paper, we present a novel analytical method for designing a single-sided plano-convex lens (PCL) capable of generating a high-quality flat-top beam in the Fresnel region from a pseudo-Gaussian THz source. The remainder of this paper is organized as follows: In Section 2, the basic necessity of beam shaping in THz imaging systems and the conventional approach for performing flat-top beam shaping are reviewed. The proposed method for single-curved flat-top beam shaping based on geometrical optics (GO) principles is introduced in Section 3, where the proposed approach integrates both collimating and beam-shaping functions into a single optical surface. The method’s efficiency is verified in Section 4 through two examples in which PCLs operating at 100 GHz and 200 GHz transform divergent Gaussian beams into flat-top profiles with target radii of 400 mm and 500 mm at working distances of 3 m and 4 m, respectively. Analytical, integral equation (IE), and finite element method (FEM) analyses confirm that the generated fields closely match those of conventional concave–convex lenses (CCLs). Sensitivity studies further demonstrate that the proposed single-curvature configuration maintains robust performance under frequency and alignment variations. Moreover, the results of this research are compared with the previously reported works. Finally, Section 5 summarizes the main findings and highlights potential applications.


\section{Conventional approach}

In quasi-optical systems, the emission from mm-wave or THz sources is commonly modeled as a pseudo-Gaussian beam \cite{Lin:10}. Under the paraxial approximation, the one-dimensional Gaussian beam intensity along the r-axis is given by: $I(r)=I_{max} \textit{exp}[-2(r/r_g)^2]$  where  $I_{max}$ is the peak intensity and $r_g$ is the $1/e^2$ waist radius, equal to half the waist diameter $\omega_0$. The performance of such a beam for illumination is characterized by two key metrics: the spill-over efficiency ($\eta_{sp}$), defined as the fraction of power falling outside the field-of-view (FoV) or aperture, and the non-uniformity ($u(r_0)$), defined as the ratio of the peak intensity $(I_{max})$ to the intensity at the edge of the FoV $(I_{0})$. High values of either metric lead to inefficient illumination and degraded imaging performance.

\begin{figure}[ht]
\centering
\includegraphics[width=\linewidth, trim={0cm 1cm 0cm 0.7cm},clip]{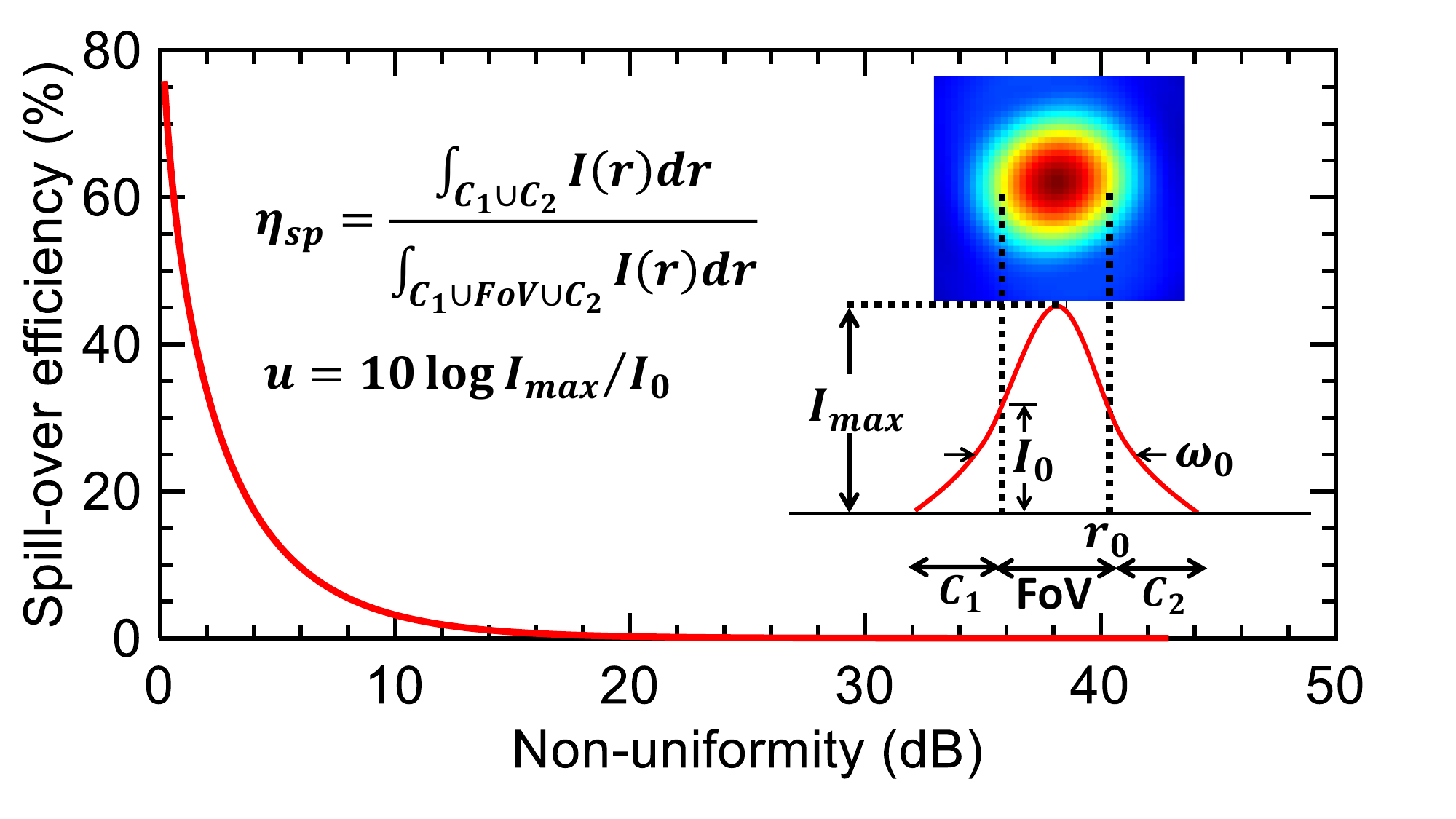}
\caption{The spill-over efficiency ($\eta_{sp}$) versus non-uniformity ($u(r_0)$) of a standard 1-D Gaussian beam in the polar coordinate system along the $r$-axis. The waist diameter of the Gaussian beam ($\omega_0$) is defined as the location where the beam intensity reaches $1/e^2$ ($13.5\%$) of its peak value ($I_{max}$) and is considered  $0.9 r_0 \leq \omega_0 \leq 13 r_0$ where $r_0$ is defined as the radius of the specific FoV.}
\label{fig_Spillover_Vs_NU}
\end{figure}

As illustrated in Fig. \ref{fig_Spillover_Vs_NU}, a fundamental trade-off exists between $\eta_{sp}$ and $u(r_0)$. Minimizing spilled power inevitably results in a brighter center relative to the edges, which compromises image quality. To resolve this conflict, beam shapers are employed in THz imaging systems to convert the Gaussian profile into a uniform flat-top intensity, enabling homogeneous object illumination \cite{abbaszadeh2020compact,cai2025high}.

\begin{figure}[ht]
\centering
\includegraphics[width=\linewidth, trim={14cm 18cm 20.5cm 10.8cm},clip]{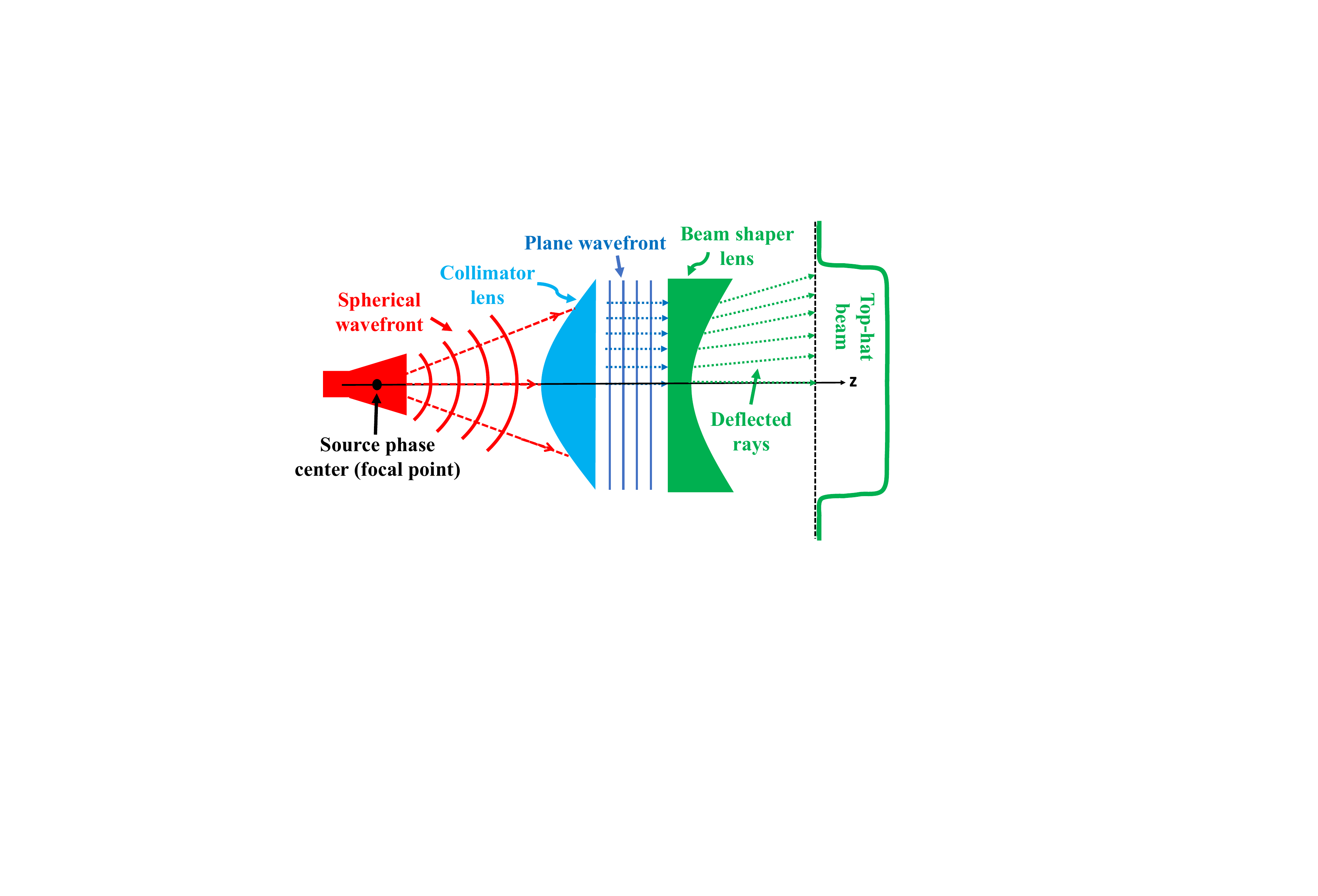}
\caption{Conventional two-lens-based flat-top beam shaping scenario. The first lens surface converts the spherical wavefront to the plane wavefront. The second lens surface converts the incident beam intensity to the flat-top beam. In practice, the two lens surfaces can be attached to form a double-sided beam shaping lens.}
\label{fig_Conventional_beam_shaping}
\end{figure}

Conventional flat-top beam shaping typically employs a two-lens system, as depicted in Fig. \ref{fig_Conventional_beam_shaping}. The first lens (collimator) transforms the divergent rays from the source into a collimated beam, while the second lens reshapes this beam to produce a flat-top intensity profile at a specific distance. Although various numerical and analytical techniques exist for designing this second beam-shaping element \cite{yu2011optimization}, they generally yield lenses with double-curved concave-convex (CCL) profiles. These designs are often bulky and inefficient for modern, compact THz imaging systems. Furthermore, the complex surface geometry significantly increases fabrication cost and complexity.

\section{Proposed method}

This work advances the state-of-the-art by introducing a novel analytical method to directly transform the spherical wavefront from a THz source into a flat-top beam at a designated Fresnel-region distance using only a single lens surface. As a proof of concept, we demonstrate the method with two practical axial beam-shaping scenarios employing the proposed PCL surface. The design is tailored for an active stand-off imaging system, analogous to the commercially available system from Terasense Inc. \cite{tzydynzhapov2020new}, which uses an IMPATT diode and conical horn antenna to radiate a coherent CW THz wave with approximately 100 mW power.

\begin{figure}[ht]
\centering
\includegraphics[width=\linewidth, trim={18.5cm 7.3cm 22.2cm 7cm},clip]{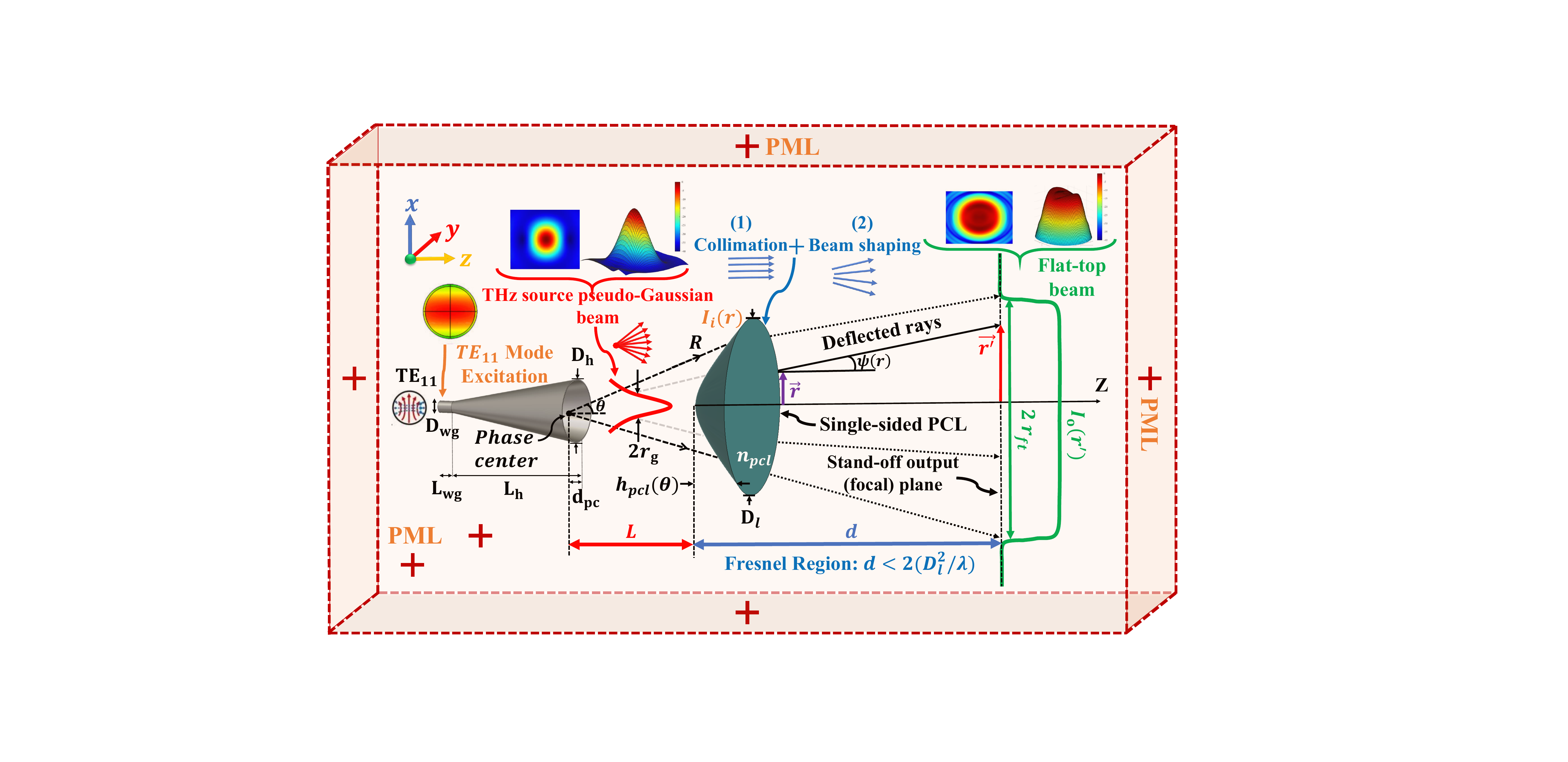}
\caption{Quasi-optical flat-top beam shaping setup utilizing the proposed PCL structure. The divergent principal (black) and chief (gray) rays from the source are collimated and diverged controllably to generate a flat-top beam shape in the Fresnel region. A perfectly matched layer (PML) is utilized to truncate the computational region. }
\label{fig_Practical_flattop_beamshaping_setup}
\end{figure}

The practical axial beam shaping setup is illustrated in Fig.  \ref{fig_Practical_flattop_beamshaping_setup}. A THz conical horn source, excited in the $TE_{11}$ dominant mode via a circular waveguide (cut-off frequency: 90.57 GHz), generates the incident spherical wavefront. The figure also presents 2-D and 3-D plots of the normalized electric field intensity, showing the transformation from the source's pseudo-Gaussian beam to the output flat-top profile.

The lens design begins with a quasi-optical approach, employing ray-optics to derive analytical formulations, with verification reserved for wave-optics or full-wave simulations. The collimator surface profile (the first lens in Fig. \ref{fig_Conventional_beam_shaping}) is derived from Fermat's principle (see supplementary information.), yielding:

\begin{equation}
R = \frac{\left(n_{pcl} - 1\right)L}{n_{pcl} \cos(\theta) - 1}
\label{eq_1}
\end{equation}

\noindent where $R$ is the distance from an arbitrary point on the lens surface to the focal point, $L$ is the focal length, $n_{PCL}$ is the refractive index of the lens, and $\theta$ is the angle relative to the optical axis (z-axis). For beam shaping, we adopt geometrical transformation techniques \cite{bryngdahl1974geometrical}. The conservation of power, $P_{in}=P_{out}$, in cylindrical coordinates is expressed as:

\begin{equation}
    \int_0^{2\pi} \int_0^r I_i(\gamma) \gamma d\gamma d\phi = \int_0^{2\pi} \int_0^{r'} I_o(\gamma') \gamma' d\gamma' d\phi
    \label{eq_2}
\end{equation}

\noindent where $r$ is the radial distance from the center of the lens aperture, $I_i$ is the incident beam intensity, and $I_o$ is the output beam intensity. For our problem, the incident Gaussian beam and the ideal output flat-top beam are defined as:

\begin{equation}
    I_i(r) = I_{max} \exp\left(-2\left(\frac{r}{r_g}\right)^2\right), \quad I_o(r') = \begin{cases} 
        1, & r' < r_{ft} \\
        0, & \text{o.w.}
    \end{cases}
\label{eq_3}
\end{equation}

Here, $r_g$ and $r_{ft}$ are the waist radius of the Gaussian beam and the radius of the output flat-top beam, respectively. Applying Eqs. (\ref{eq_2}) and (\ref{eq_3}) to the rotationally-symmetric geometry in Fig. \ref{fig_Practical_flattop_beamshaping_setup} gives: 

\begin{equation}
    \int_{0}^{r} I_i(\gamma) \gamma d\gamma = \left(\frac{r'}{r_{ft}}\right)^2 \int_{0}^{D_l/2} I_i(\gamma) \gamma d\gamma
\label{eq_4}
\end{equation}

\noindent where $D_l$ is the diameter of the lens. The beam shaping phase profile $\phi_b$ is governed by the generalized Fermat's principle \cite{yu2011light}:

\begin{equation}
\frac{d\phi_b}{dr}=k\sin(\psi(r))
\label{eq_5}
\end{equation}

\noindent where $k=2\pi/\lambda$ is the free space wavenumber, $\lambda$ is the operating wavelength, and $\psi(r)$ is the deflection angle. Under the paraxial approximation, Eq. (\ref{eq_5}) simplifies to $d\phi_b/dr=k\tan(\psi(r))$. From Fig. \ref{fig_Practical_flattop_beamshaping_setup}, $\tan(\psi(r))=(r'-r)/d$, where $d$ is the distance to the output plane. Substituting this and Eq. (\ref{eq_4}) into the phase profile relation yields:

\begin{equation}
\frac{d\phi_b}{dr} = \frac{2\pi}{\lambda d} \left\{ r_{ft} 
\sqrt{ \frac{\int_{0}^{r} I_{i}(\gamma)\,\gamma \, d\gamma}
{\int_{0}^{D_l/2} I_{i}(\gamma)\,\gamma \, d\gamma} } - r \right\} 
\label{eq_6}
\end{equation}

Integrating both sides of this differential equation and substituting the Gaussian intensity from Eq. (\ref{eq_3}) results in the phase profile:

\begin{equation}
\phi_b = U + \frac{2 \pi}{\lambda d} 
\int_{0}^{r} \Bigg\{ r_{ft} 
\left[ \sqrt{\frac{1-\exp\left(-2\left(\tfrac{\gamma}{r_g}\right)^2\right)}
                {1-\exp\left(-2\left(\tfrac{D_l}{2 r_g}\right)^2\right)}} \right] - \gamma \Bigg\} \, d\gamma
, \quad 0 < r < \tfrac{D_l}{2}
\label{eq_7}
\end{equation}

\noindent where $U$ is an integration constant. Using the thin element approximation, $\Delta\phi=k\Delta n_{pcl}h(r)$ \cite{schmidt2016wave}, the beam shaper height profile $h_{bs}(r)$ is obtained from Eq. (\ref{eq_8}):

\begin{equation}
h_{bs}(r) = h_0 + \frac{1}{(n_{pcl}-1)d} 
\int_{0}^{r} \Bigg\{ r_{ft} 
\left[ \sqrt{\frac{1-\exp\!\left(-2\left(\tfrac{\gamma}{r_g}\right)^2\right)}
                {1-\exp\!\left(-2\left(\tfrac{D_l}{2 r_g}\right)^2\right)}} \right] \\
- \gamma \Bigg\} \, d\gamma
\label{eq_8}
\end{equation}

Assuming $h_0=0$, the final PCL profile, which combines collimation and beam shaping, is given by:

\begin{equation}
    h_{pcl}(\theta) = \alpha_{opt} h_{bs}(r) + R(\theta) \cos(\theta)-L
\label{eq_9}
\end{equation}

Here, $\alpha_{opt}$ is an optimum scaling factor determined through full-wave simulation and $r=R(\theta)\sin(\theta)$. The lens surface is generated by revolving this profile about the z-axis ($\theta=0$). This formulation enables the single PCL surface to perform simultaneous beam collimation and flat-top beam shaping.


\section{Results and discussions}

Following the proposed method, we designed two rotationally symmetric PCLs for distinct Fresnel-region beam-shaping scenarios, as illustrated in Fig. \ref{fig_Practical_flattop_beamshaping_setup}. The design employs a THz conical horn source with the following specifications: $d_{pc}$=9 mm, $D_h$=12 mm, $D_{wg}$=1.9 mm, $L_h$=27 mm, $L_{wg}$=3 mm, $r_g$=11.9 mm. The parameters for the first scenario are: operating wavelength $\lambda$=3 mm, antenna-to-lens distance L=52 mm, lens diameter $D_l$=100 mm, output plane distance d=3 m, lens refractive index $n_{pcl}$= 1.45, target flat-top beam radius $r_{ft}$=400 mm, and optimum scaling factor $\alpha_{opt}$=0.79. The second scenario uses identical parameters except for $\lambda$=1.5 mm, d=4 m, $r_{ft}$=500 mm, and $\alpha_{opt}$=1. The resulting lens thicknesses are 42.11 mm and 41.54 mm for the first and second designs, respectively. For comparison, a conventional double-curvature CCL \cite{abbaszadeh2019generating} was also designed for the first example. In this case, the same design parameters used for the CCL are adopted to explicitly justify and demonstrate that the proposed PCL generates a comparable flat-top beam with similar characteristics. However, in the second example, different values are selected for the working distance $d$ and the target flat-top beam radius $r_{ft}$ to show that the proposed analytical approach remains effective and reliable regardless of the operating scale. It is worthy of pointing out that the PCL is intended to generate a flat-top beam within its Fresnel region, which is suitable for active stand-off imaging systems where sufficiently large working distances and FoV are required.

\begin{figure}[ht]
\centering
\includegraphics[width=10cm, trim={26cm 15.2cm 27cm 11.2cm},clip]{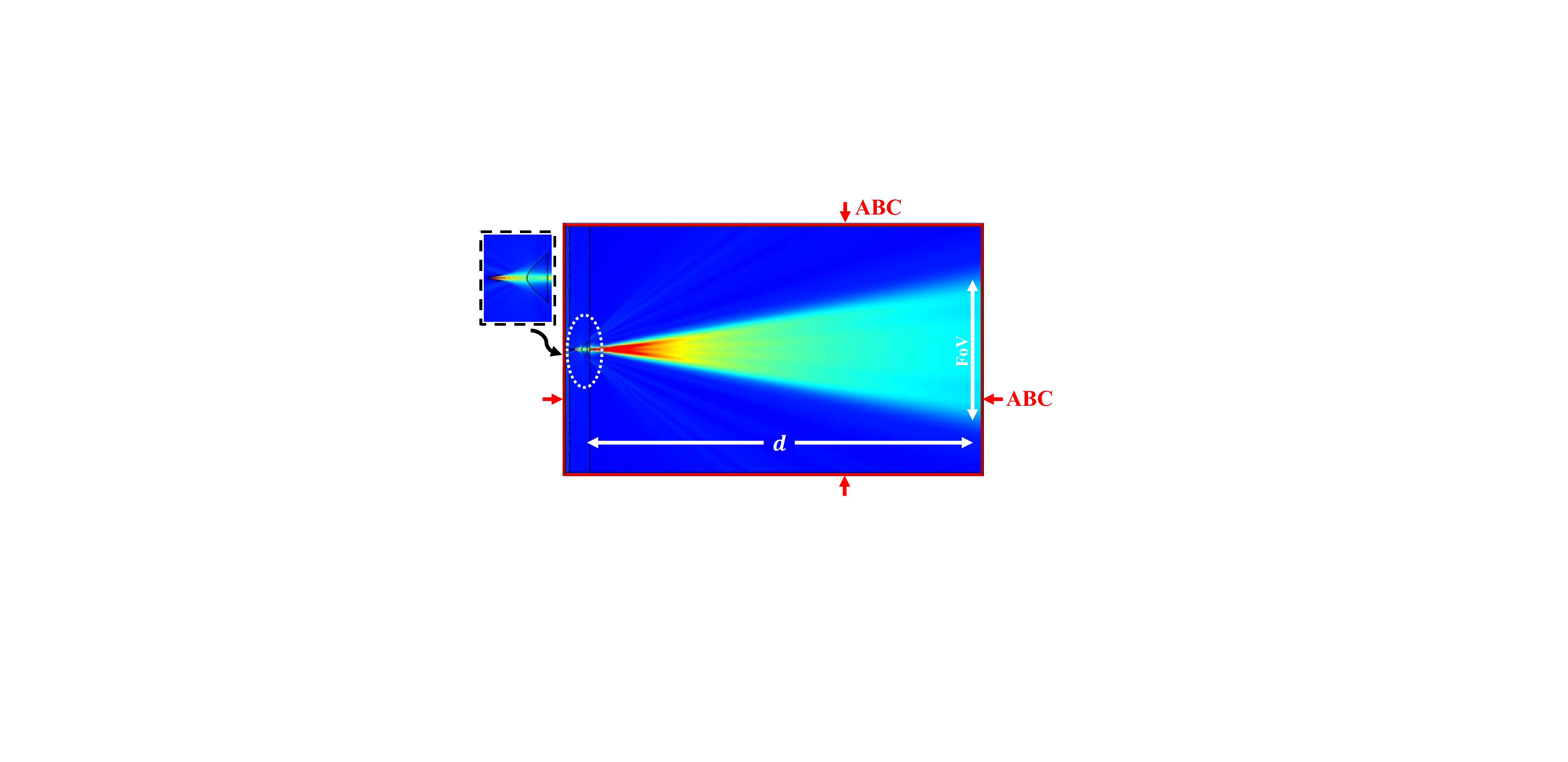}
\caption{The FEM simulation setup.The PCL is excited by the horn antenna and homogeneously illuminates the FoV at the working distance of $d$ .}
\label{fig_2D_FEM_Setup}
\end{figure}

Given the large electrical size of the structures, simulations were performed using an integral equation solver with the multi-level fast multipole method (MLFMM) for computational efficiency. The results were verified using both the analytical approach and a FEM analysis, the latter shown in Fig. \ref{fig_2D_FEM_Setup}. In Fig. \ref{fig_2D_FEM_Setup}, absorbing boundary conditions (ABC) are used, and the longitudinal evolution of the electric field distribution visualizes the transition from a quasi-Gaussian to a uniform flat-top beam.

\begin{figure}[ht]
\centering
\includegraphics[width=\linewidth, trim={25cm 32.5cm 29cm 6.3cm},clip]{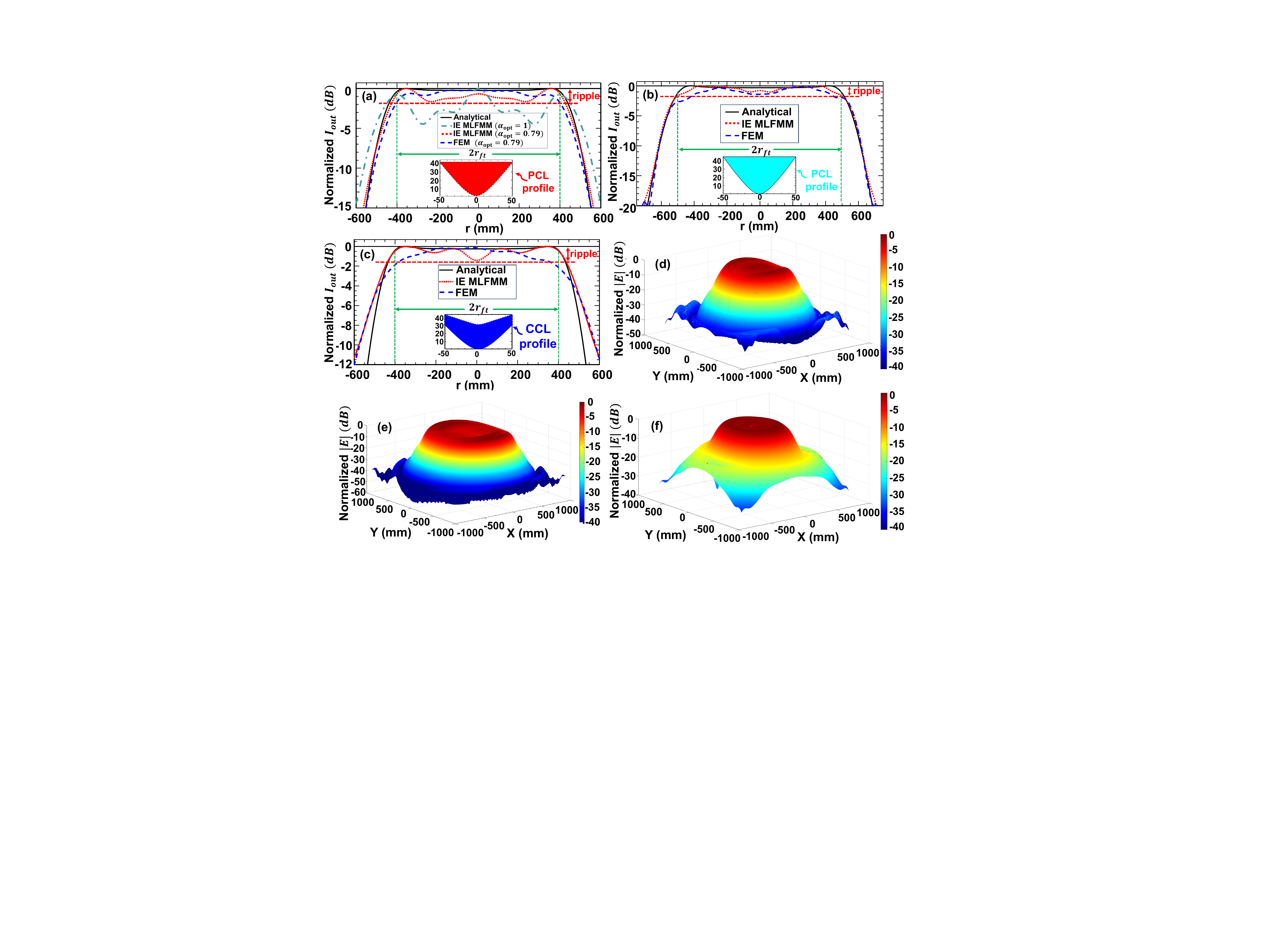}
\caption{The extracted PCL height profile and the calculated output flat-top beam with analytical method, IE method, and FEM for the first and second examples are shown in (a) and (b), respectively. Similarly, the results for the CCL surface, calculated for the first example, are illustrated in (c). The simulated 3D output flat-top beam of the PCL using the IE method at $100 GHz$ and $200 GHz$ is shown in (d) and (e), respectively, and the result for the CCL surface at $100 GHz$ is shown in (f).}
\label{fig_beamshaping_results}
\end{figure}

The designed PCL profiles and the resulting flat-top beam intensities for both examples are presented in Fig. \ref{fig_beamshaping_results}. The analytical results are derived at the distance of $d$ using the Fresnel diffraction integral\cite{Goodman2004}. Beam quality was quantified using ripple (uniformity) and roll-off (edge sharpness, defined as the ratio of full width at -3 dB to full width at -10 dB). For the first optimized PCL (Fig. \ref{fig_beamshaping_results}(a)), the ripple is $\pm0.8$ dB and the roll-off is -0.7. The corresponding output intensity of the optimized PCL demonstrates improved uniformity compared to the initial design, as its ripple ( $\pm2.1$ dB in the preliminary PCL profile)  is substantially lower. This clearly highlights the necessity of numerical optimization. The second PCL (Fig. \ref{fig_beamshaping_results}(b)) performs similarly well, with a ripple of $\pm0.5$ dB and a roll-off of -0.6. These values confirm the high quality of the generated flat-top beams. Moreover, they indicate that additional numerical refinement is unnecessary in the second example, since the output intensity demonstrates superior uniformity with sharper roll-off compared to the optimized PCL in the first example.   Excellent agreement is observed between the analytical and numerical results.

For reference, the CCL result for the first scenario (Fig. \ref{fig_beamshaping_results}(c)) shows a comparable ripple of $\pm0.7$ dB and a roll-off of -0.91. Crucially, however, the proposed PCL achieves this performance with a significantly simpler plano-convex profile, which offers substantial advantages in fabrication ease and cost. The uniform illumination capability of the PCL designs is further demonstrated by the 3D output beams calculated using the IE method, shown for both PCL examples (Figs. \ref{fig_beamshaping_results}(d)-(e)) and the CCL (Fig. \ref{fig_beamshaping_results}(f)). 

\begin{figure}[ht]
\centering
\includegraphics[width=\linewidth, trim={2.5cm 5cm 3cm 2cm},clip]{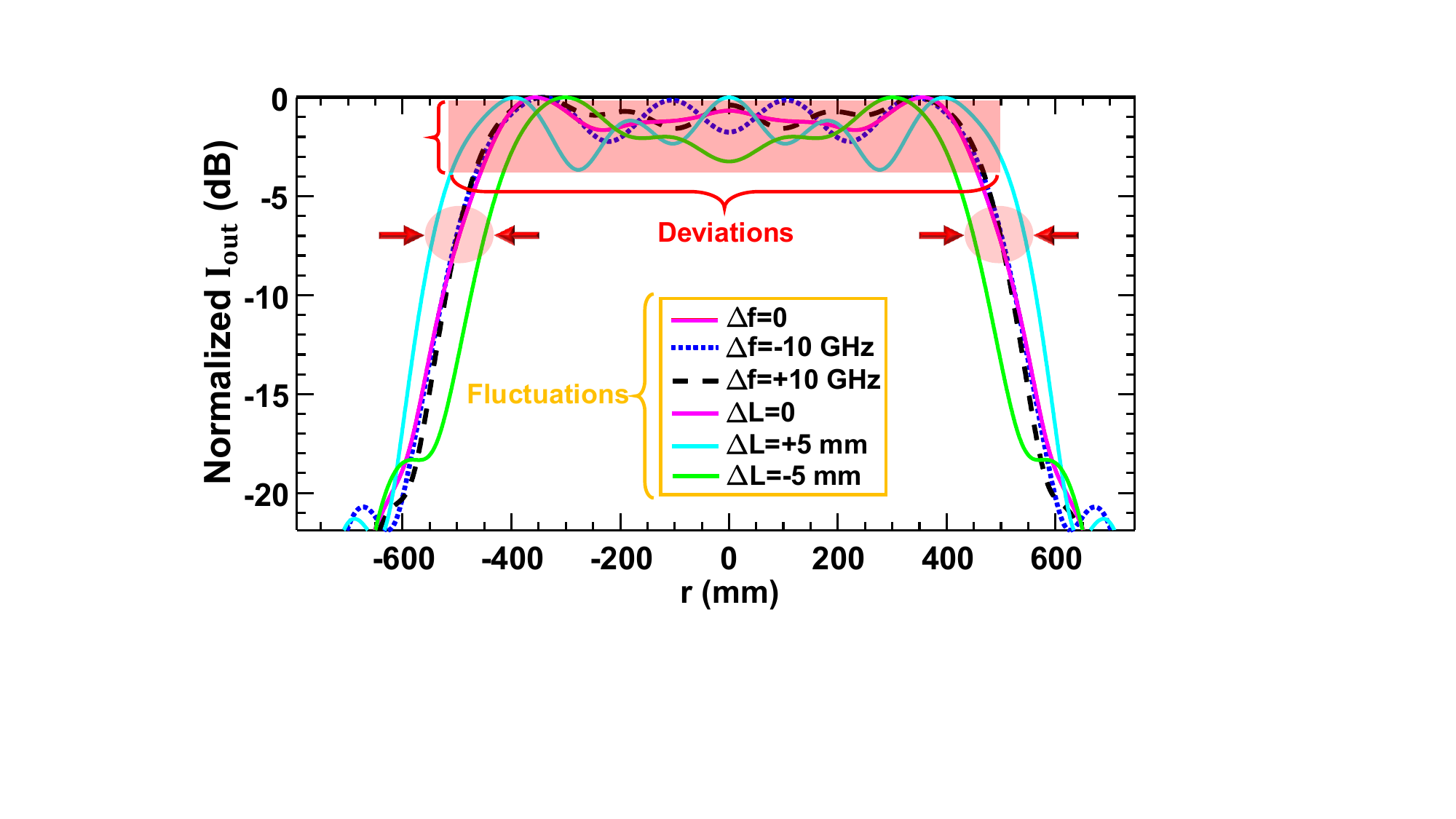}
\caption{Sensitivity of the output flat-top beam to the deviation in frequency ($\Delta f$) and source-lens distance ($\Delta L$) of the first example.}
\label{fig_sensitivity_analysis}
\end{figure}

Also, Figure \ref{fig_sensitivity_analysis} presents a sensitivity analysis of the 100 GHz PCL design, where a $\pm10\%$ change in frequency has a minimal impact on the beam radius and roll-off. The ripple, however, is more sensitive, increasing to $\pm1.08$ dB at 90 GHz. Furthermore, a $\pm5$ cm variation in L has a negligible effect on the roll-off, while the beam radius changes by approximately $\Delta r_{ft}/r_{ft} \cong \pm11 \%$. This behavior can be understood by analyzing the phase profile. When the lens is displaced from its design position ($L$=52 mm) to $L + \Delta L$, the total phase on the PCL aperture is approximated by:

\begin{equation}
    \varphi_{\text{tot}}(r) \approx \varphi_b(r) - k_0(L + \Delta L) + \frac{k_0}{2} \left(\frac{r}{L}\right)^2 \Delta L
\label{eq_11}
\end{equation}

The final term in Eq. (\ref{eq_11}) acts as a diverging lens when $\Delta L$ > 0, explaining the observed beam widening. A comparative analysis with prior works is summarized in Table \ref{Table_I}, evaluating key metrics such as bandwidth, efficiency, roll-off ($\alpha$), ripple ($\Delta r$), and fabrication complexity. The plano-convex lens is easiest to manufacture with a single curved surface, while concave-convex lenses require expensive, dual-surface precision, and metasurface lenses necessitate complex nanofabrication. Moreover, it holds a superior position in compact imaging systems involving dielectric lenses, as it enables beam shaping with just one lens rather than with a pair of aspheric lenses or a system comprising a separate collimating lens and a Powell beam-shaping lens.

\begin{table}[htbp]
\centering
\caption{\bf Comparison of the beam shaping results of this work with the reported works.}
\renewcommand{\arraystretch}{1.25}
\small
\setlength{\tabcolsep}{6pt}

\resizebox{\textwidth}{!}{
\begin{tabular}{
|>{\centering\arraybackslash}p{1.8cm}|   
 >{\centering\arraybackslash}p{3.6cm}|   
 >{\centering\arraybackslash}p{2.4cm}|
 >{\centering\arraybackslash}p{2cm}|
 >{\centering\arraybackslash}p{2.4cm}|
 >{\centering\arraybackslash}p{2cm}|
 >{\centering\arraybackslash}p{2cm}|
}
\hline
\textbf{Reference} & \textbf{Method} & \textbf{BW (THz)} & \boldmath $\eta$ & \boldmath $\alpha$ & \boldmath $\Delta r$ & \textbf{Fabrication} \\
\hline

\textbf{\cite{zhang2025terahertz}} & 3-layer metallic metasurface lens & 0.4–1.1 & 78.03\% & – & $u(x_0)=0.126$ & Complex \\
\hline

\textbf{\cite{abbaszadeh2019generating}} & Double curved dielectric lens & 0.1 & 86.3\% & -0.91 & $\pm0.7$ dB & Moderate \\
\hline

\textbf{\cite{price2020subterahertz}} & Two plano-convex aspheric lenses & 0.95-0.3 & 70$~$75$\%$ & - & 14\% & Moderate \\
\hline

\textbf{\cite{bezziou2023design}} & Powell Lens & 0.85-0.115  & 90.97\% & -0.43 & $\pm$
10–15\% & Moderate \\
\hline

\textbf{\cite{abbaszadeh2020compact}} & All-dielectric silicon metasurface lens & 0.1 & 47.4\% & - & - & Complex \\
\hline

\textbf{This work} & Single curved dielectric lens & 0.1–0.2 & 72.3\% & -0.7 & $\pm0.8$ dB & Easy \\
\hline

\end{tabular}
}

\label{Table_I}
\end{table}


\section{Conclusion}

In this work, a simple and efficient mathematical method is proposed to perform Gaussian to flat-top beam shaping in the Fresnel region. Two single-sided PCLs are designed to shape the pseudo-Gaussian beam of 100 GHz and 200 GHz into a flat-top intensity at a distance of 3 m and 4 m, respectively. Analytical method, as well as IE and FEM methods, are used to calculate the output beam, and the obtained results are compared with a double-sided CCL. Furthermore, the sensitivity of the proposed PCL with respect to deviations in frequency and source-lens distance is investigated. The proposed method simplifies fabrication, enables more compact lens designs, and is well-suited for beam shaping in the state-of-the-art THz imaging systems.

\section*{Disclosures}
The authors declare no conflicts of interest.

\section*{Data Availability Statement}
Data underlying the results presented in this paper are not publicly available at this time but may be obtained from the authors upon reasonable request.

\section*{Supplemental document}
See Supplement 1 for supporting derivations.


\bibliographystyle{unsrt}
\bibliography{references}






\end{document}